\documentclass[12pt]{article}
\usepackage{amstex}

\begin{document}
\vspace*{-.6in}
\thispagestyle{empty}
\begin{flushright}
CALT-68-2099\\
hep-th/9702133
\end{flushright}
\baselineskip = 20pt

\vspace{.5in}
{\Large
\begin{center}
Dual D-Brane Actions\footnote{Work
supported in part by the U.S. Dept. of Energy under Grant No.
DE-FG03-92-ER40701.}
\end{center}}

\begin{center}
Mina Aganagic, Jaemo Park, Costin Popescu, and John H. Schwarz\\
\emph{California Institute of Technology, Pasadena, CA  91125, USA}
\end{center}
\vspace{1in}

\begin{center}
\textbf{Abstract}
\end{center}
\begin{quotation}
\noindent Dual super Dp-brane actions are constructed by carrying out a duality
transformation of the world-volume $U(1)$ gauge field. The resulting world-volume actions,
which contain a $(p-2)$-form gauge field, are shown to have the expected
properties. Specifically, the D1-brane and D3-brane transform in ways
that can be understood on the basis of the $SL(2,Z)$ duality of type IIB superstring
theory. Also, the D2-brane and the D4-brane transform in ways that are
expected on the basis of the relationship between type IIA superstring theory
and 11d M theory. For example,  the dual  D4-brane action is shown to
coincide with the double-dimensional reduction of the recently constructed
M5-brane action. The implications for gauge-fixed D-brane actions are
discussed briefly. 
\end{quotation}
\vfil

\newpage

\pagenumbering{arabic}

\section{Introduction}

Several groups have recently constructed
supersymmetric  D-brane actions with local kappa symmetry~\cite{cederwall1,aganagic,bergshoeff1}. 
These supersymmetric actions are a good 
starting point for studying various properties of D-branes. For example,
since D-branes play a reasonably well understood role in the web of string dualities, 
there are definite expectations for how 
each D-brane action  should transform
under a duality transformation
of its world-volume gauge field. These reflect the $SL(2,Z)$ S duality
of the type IIB superstring theory~\cite{hull} 
and the relationship between type IIA superstring theory at strong coupling
and 11d M theory compactified on a circle~\cite{townsenda,witten10}.
Previous studies of these duality transformations have been carried out 
in the context of the bosonic truncation of the D-brane 
action~\cite{schmidhuber,tseytlin}. 
Since the super D-branes are BPS objects, 
it is appropriate to study their duality properties including the
fermionic degrees of freedom. 
Such an investigation, which is possible now that the super D-brane actions
are known, is the purpose of this paper. 

In ref.~\cite{aganagic} we only considered super D-branes in a 
flat background. Here, as a modest extension of this,
we include a constant dilaton background
for the type IIA D-branes and constant dilaton and axion backgrounds for the 
type IIB D-branes. 
Starting with the D1-brane, we show that one can obtain the 
expected $SL(2,Z)$ multiplet
of type IIB strings~\cite{schwarz10,witten11},
with the correct tensions, by performing duality transformations. 
In the case of the D2-brane, we show that the dual action describes 
the M2-brane with one target space dimension compactified. 
(The relationship between the D2-brane and the M2-brane has been 
discussed previously~\cite{townsend}, so this part is mostly review.)
In particular, we verify that the dilaton 
dependence of the D2-brane correctly reproduces the relation between the 
string metric of the IIA theory and the 11d metric of the 
M theory. This implies that the type IIA string coupling constant is correctly related
to the radius of the 11th dimension~\cite{witten10}.
Next we show that the D3-brane action is mapped into an equivalent 
D3-brane action by the duality transformation, thereby verifying 
the expected $SL(2,Z)$ invariance of the D3-brane. 
Another correspondence suggested by the duality between
M theory and type IIA superstring theory is that the double-dimensional 
reduction of the M5-brane action should coincide with the duality
transformed D4-brane action. This could not be checked previously, since we 
did not have a suitable supersymmetric M5-brane action. 
However this action has been constructed recently~\cite{APPS}, 
so we are now in a position to verify that 
the dual D4-brane action is identical to the double-dimensional 
reduction of the M5-brane action as expected. (The M5-brane has also been discussed
recently in refs.~\cite{pasti,howe}.)
Finally, we indicate how 
duality transformations relate specific gauge choices for the gauge-fixed D-brane actions. 
 
The calculation of the duality transformations of supersymmetric D-brane actions
is quite similar to that  of the bosonic 
actions described previously 
in refs.~\cite{schmidhuber,tseytlin,dealwis,lozano}. 
Since the behavior of the 
fermionic degrees of freedom under the duality transformation
is the new ingredient, this is the part of the analysis that is
emphasized.  Unless otherwise stated,
the conventions used here are the same as those of ref.~\cite{aganagic}.

\medskip  
 \section{Dual Born--Infeld Actions}

The essential steps involved in world-volume duality transformations 
of D-brane actions can be described for
the simpler problem of Born--Infeld theory.  Subsequent sections will discuss
the extensions that are required for various supersymmetric D-brane actions.
Born--Infeld theory in $n = p + 1$ dimensions is given by
\begin{equation}\label{A}
S = - \int d^n \sigma \sqrt{-\det (\eta_{\mu\nu} + F_{\mu\nu})},
\end{equation}
where $\eta_{\mu\nu}$ is the Minkowski metric and $F_{\mu\nu} = \partial_\mu
A_\nu - \partial_\nu A_\mu$ is the Maxwell field strength.  The basic idea is
to recast the theory in terms of a dual $(p - 2)$-form potential
$B_{\mu_{1}\mu_{2}\ldots \mu_{p - 2}}$ given by
\begin{equation}\label{B}
- {\delta S\over\delta F_{\mu\nu}} =  \tilde{H}^{\mu\nu} = {1\over (p -
2)!} \epsilon^{\mu\nu\lambda\rho_{1}\ldots\rho_{p-2}} 
\partial_{\lambda}B_{\rho_{1}\ldots \rho_{p-2}}.
\end{equation}

The Bianchi identity for the $B$ field is the field equation for the Maxwell
field.  Also, the Bianchi identity of the Maxwell field provides the field
equation for the $B$ field.  To make the latter equation explicit one needs to solve
eq.~(\ref{B}) for $F_{\mu\nu}$.  Then one can construct an action that gives
the field equation.  Equivalently, one can add a Lagrange multiplier term
${1\over 2}\tilde{H}^{\mu\nu} (F_{\mu\nu} - 2\partial_\mu A_\nu)$ to eq.~(\ref{A})
and eliminate $F$.

To solve eq.~(\ref{B}) for $F_{\mu\nu}$, it is convenient to use Lorentz invariance
to bring $F_{\mu\nu}$ to the  canonical form
\begin{equation}
F_{\mu\nu} = \left(\begin{matrix}
\,0 \quad \, f_1 && \\
-f_1\quad  0  &&\\
&\, \,0\quad f_2 &\\
&-f_2 \quad 0 &\\
& & \ddots\\
\end{matrix}\right).
\end{equation}
Then eq.~(\ref{B}) implies that $\tilde{H}^{\mu\nu}$ has the same structure
\begin{equation}
\tilde{H}^{\mu\nu} = \left(\begin{matrix}
\,0 \quad \, h_1 && \\
-h_1\quad  0  &&\\
&\, \,0\quad h_2 &\\
&-h_2 \quad 0 &\\
& & \ddots\\
\end{matrix} \right).
\end{equation}
In this notation, eq.~(\ref{B}) becomes\footnote{This is the formula for Euclidean signature.
The extension to Lorentzian signature is straightforward.}
\begin{equation}
h_i =  {f_i\over 1 + f_i^2} \sqrt{  \prod (1 + f_j^2)}.
\end{equation}
For $p \leq 4$ there are at most two $f$'s and this equation can be inverted.
The result for this case is
\begin{equation}
f_i =  {h_i\over 1 - h_i^2} \sqrt{ \prod (1 - h_j^2)}.
\end{equation}
Unfortunately, for $p > 4$, when there are three or more $f$'s, we are unable to
carry out the inversion.  With three $f$'s we find a quintic equation. 
The coefficients of the quintic are not completely generic, so a closed form
solution might exist, but we have not found one.

Having found the field equation of the $B$ field in a special frame, it is easy
to pass to a general frame and write an action that gives the desired equation.
The result is (the subscript $D$ stands for ``dual'')
\begin{equation}
S_D = - \int d^n \sigma \sqrt{- \det (\eta_{\mu\nu} + i \tilde{H}_{\mu\nu})}.
\end{equation}
We emphasize, once again, that this result is only correct for $p \leq 4$ or $n \leq 5$.

Now consider the more general Born--Infeld action
\begin{equation}
S = - \int d^n \sigma \sqrt{ - \det (G_{\mu\nu} + {\cal F}_{\mu\nu})},
\end{equation}
where $G_{\mu\nu}$ is a symmetric tensor field and ${\cal F}_{\mu\nu} =
F_{\mu\nu} - b_{\mu\nu}$ is an antisymmetric tensor field.  Repeating the
analysis described above in this more general setting gives (for $p \leq 4)$
\begin{equation}
S_D = - \int d^n \sigma (\sqrt{-\det (G_{\mu\nu} + i K_{\mu\nu})} + {1\over 2}
\tilde{H}^{\mu\nu} b_{\mu\nu}), \label{moregen}
\end{equation}
where
\begin{equation}
K_{\mu\nu} = {1\over\sqrt{-\det G}} G_{\mu\rho} G_{\nu\lambda}
\tilde{H}^{\rho\lambda}.
\end{equation}
The $H \wedge b_2$ term in $S_D$ will be identified as part of the Wess--Zumino term of
the dual D-brane.

The analysis described here is not the whole story for super D-branes, since
they also contain Wess--Zumino terms that are polynomial functions of
$F_{\mu\nu}$.  Specifically, they are linear in $F$ for $p = 2,3,$ quadratic in
$F$ for $p = 4,5$, and so forth.  The extension of the analysis given here to include
these terms will be described on a case-by-case basis in the sections that
follow.

\medskip
\section{The D1-brane}

Let us consider the D1-brane ({\it i.e.}, the type IIB D-string) first. If we include the
dependence on a {\it constant} dilaton $\phi$, the action of the super D1-brane with 
kappa symmetry is given by 
\begin{equation}
S = \int d^2 \sigma \Big( -e^{-\phi} \sqrt{-{\rm det}\, (G_{\mu\nu}+\cal F_{\mu\nu})} 
+ L_{WZ} \Big). 
\label{eq:sd1}
\end{equation}
Here $G_{\mu\nu} =  \eta_{mn} \Pi_{\mu}^m \Pi_{\nu}^n$, where
$\Pi^m_{\mu} = \partial_{\mu}X^m -\bar\theta \Gamma^m \partial_{\mu} \theta$.
Also, $X^m$ and $\theta$ are coordinates of type IIB superspace and $\eta_{mn}$ is the 
10d Minkowski metric. The induced world volume metric $G_{\mu\nu}$
is the supersymmetrized pullback of the 10d string metric $ \eta_{mn}$.
Also, ${\cal F}=F-b_2$ with $b_2=-\bar{\theta}\tau_3 \Gamma_m d\theta 
(dX^m +\frac{1}{2}\bar{\theta}\Gamma^m d\theta)$.
$L_{WZ}$ denotes the Wess--Zumino term, which can be represented 
by a 2-form on the world volume of the D1-brane. 
Specifically, 
\begin{equation}
S_{WZ} = \int d^2 \sigma L_{WZ} = e^{-\phi} \int_{M_2} C_2, \label{eq:wz1}
\end{equation}
where $C_2=\bar{\theta}\tau_1 \Gamma_m d\theta (dX^m +\frac{1}{2}
\bar{\theta}\Gamma^m d\theta)$ and $dC_2=d\bar{\theta}\tau_{1}
\Gamma_m d\theta \Pi^m$ with $\Pi^m=dX^m+\bar{\theta}\Gamma^m d\theta$.
Note that eq.~(\ref{eq:sd1}) is the D1-brane  
action of ref.~\cite{aganagic} rescaled by the string coupling constant. 
One can also add a total derivative term (analogous to the $\theta$ term
of QCD) to the Wess--Zumino term in 
eq.~(\ref{eq:wz1}):
\begin{equation}
e^{-\phi} C_2 \rightarrow e^{-\phi} C_2 -C_0 F,
\end{equation}  
where $C_0$ is a constant ``axion'' background field.
Since $C_0 F$ is a total derivative, it does not affect the classical equations 
of motion. A constant shift of $C_0$ is a trivial classical symmetry of
the action~(\ref{eq:sd1}). In the quantum theory it is replaced by a quantized
shift, just as in QCD. This reflects the breaking of the classical $SL(2,R)$ symmetry
to the quantum $SL(2,Z)$ symmetry.

Now let us perform the duality transformation. Following ref.~\cite{tseytlin},
one introduces a Lagrange multiplier field
$\tilde{H}^{\mu\nu}=-\tilde{H}^{\nu\mu}$ as follows
\begin{equation}
S^{\prime}= \int d^2 \sigma \Big(- e^{-\phi} \sqrt{-{\rm det}\, (G_{\mu\nu}+
{\cal F}_{\mu\nu})}
+\frac{1}{2} \tilde{H}^{\mu\nu} (F_{\mu\nu} - 2 \partial_{\mu} A_{\nu})
+  \frac{1}{2} e^{-\phi}\epsilon^{\mu\nu}C_{\mu\nu} 
-\frac{1}{2}C_0 \epsilon^{\mu\nu} F_{\mu\nu} \Big)
 \label{eq:sd2} 
\end{equation}
and considers $F_{\mu\nu}$ to be an independent 
field.
%\footnote{As explained in \cite{tseytlin}, 
%$L^{\prime}(F, \tilde{A})=L(F)+F d\tilde{A}$ is equivalent to $L(dA)$.}
Varying $A_{\nu}$ gives  $\partial_{\mu}\tilde{H}^{\mu\nu}=0$, 
which implies that $\tilde{ H}^{\mu\nu}=\epsilon^{\mu\nu}
\Lambda$ with $\Lambda$ constant.
This gives $S^{\prime}=S_{1}+S_{2}$, where  
\begin{equation}
S_{1}= \int d^2 \sigma \Big( -  e^{-\phi} \sqrt{-{\rm det}\, (G_{\mu\nu}+
{\cal F}_{\mu\nu})}
+\frac{1}{2} (\Lambda- C_0 )\epsilon^{\mu\nu} {\cal F}_{\mu\nu}\Big)
\end{equation}
\begin{equation}
S_{2}=\int e^{-\phi} C_2 + (\Lambda -C_0) b_2. \label{eq:s2}
\end{equation}
Our convention is that whenever an integral appears
without a $d^n \sigma$ it is an 
integral of a differential form. It can be easily converted to a
usual integral. For example, $\int F
=\int d^2 \sigma \frac{1}{2} \epsilon^{\mu\nu} F_{\mu\nu}$. 
The basic strategy, described in the preceding section,
is to use the equation of motion for $F$ to rewrite  
the action in terms of $\Lambda$ instead of $F$. 
The duality transformation of $S_1$ is the same as the bosonic case, 
if we replace $F$ by $\cal F$. Thus the dual of $S_1$ is
\begin{equation}
S_{1D}= -\int d^2 \sigma \sqrt{e^{-2\phi}+(\Lambda-C_0)^2}
\sqrt{-{\rm det}\, G_{\mu\nu}},
\end{equation}
while $S_{2}$ is unaffected by the duality transformation. 

In eq.~(\ref{eq:s2}), we have 
\begin{equation}
e^{-\phi} C_2 +(\Lambda-C_0) b_2=\bar{\theta}
(e^{-\phi}\tau_{1}-(\Lambda-C_0)\tau_3)\Gamma_m d\theta 
\wedge (dX^m+\frac{1}{2}\bar{\theta}
\Gamma^m d\theta).
\end{equation}
Since the eigenvalues of 
$e^{-\phi}\tau_{1}-(\Lambda-C_0)\tau_3$ are 
$\pm \sqrt{e^{-2\phi}+(\Lambda-C_0)^2}$ we can redefine 
\begin{equation}
 e^{-\phi}\tau_{1}-(\Lambda-C_0)\tau_3 \equiv 
\sqrt{e^{-2\phi}+(\Lambda-C_0)^2} \tau_3^{\prime}. 
\end{equation}
Then the total action can be written as 
\begin{equation}
S_D= \sqrt{e^{-2\phi}+(\Lambda-C_0)^2}
\int d^2 \sigma \Big(-\sqrt{-{\rm det}\, G_{\mu\nu}}
-\frac{1}{2}\epsilon^{\mu\nu}b_{\mu\nu}^{\prime} \Big)  \label{eq:S1d} 
 \end{equation}
with $b_2^{\prime}=-\bar{\theta}\tau_3^{\prime} 
\Gamma_m d\theta \wedge (dX^m +\frac{1}{2}
\bar{\theta}\Gamma^m d\theta)$. This is nothing but the kappa-symmetric 
superstring action with the modified tension 
\begin{equation}
T^{\prime}=\sqrt{e^{-2\phi}+(\Lambda-C_0)^2}. \label{eq:tension}
\end{equation}
This agrees with the tension formula derived in ref.~\cite{schwarz10}
for the $SL(2,Z)$ covariant spectrum of 
strings provided that one identifies the integer value $\Lambda = m$ 
as corresponding to the 
 $(m,1)$ string in a background with  constant dilaton $\phi$
and axion $C_0$. An 
equivalent interpretation is that eq.~(\ref{eq:S1d}) describes the fundamental $(1,0)$ 
string with an $SL(2,Z)$ transformed metric, dilaton and axion. 
(The canonical Einstein metric is invariant, but the string metric is not.)
The relevant $SL(2,Z)$ transformations maps $C_0+ie^{-\phi}$ to 
$-(C_0-\Lambda+ie^{-\phi})^{-1}$. Thus the coupling constant of the 
fundamental string after the duality transformation is given by
$e^{\tilde{\phi}}=e^{-\phi}+e^{\phi}(\Lambda-C_0)^2$. 
%Indeed, if we define 
%$\sqrt{e^{-2\phi}+(\Lambda-C_0)^2}\sqrt{-{\rm det}\, G_{\mu\nu}}\equiv
%\sqrt{-{\rm det}\, G^{\prime}_{\mu\nu}}$, 
%then $G^{\prime}_{\mu\nu}= \sqrt{(e^{-2\phi}
%+(\Lambda-C_0)^2)}G_{\mu\nu}$, which 
%can be written as $e^{\tilde\phi/2}g_{\mu\nu}$,
%where $g_{\mu\nu}$ is $SL(2,R)$-invariant Einstein metric. 

\medskip

\section{The D2-brane}

The D2-brane action was the first of the super D-brane actions to be worked out.
The method that was used was to start from the known M2-brane action~\cite{bergshoeff1}
and to perform a duality transformation of a world volume scalar field
corresponding to a circular target-space coordinate~\cite{townsend}. The dual of 
a scalar in 3d is a $U(1)$ gauge field, of course.
Here we reverse the argument, starting from the D2-brane action to get the 
M2-brane action. 
Consider the super D2-brane action in the string metric
\begin{equation}
S = \int d^3 \sigma \Big( - e^{-\phi} \sqrt{-{\rm det}\,
(G_{\mu\nu}+\cal F_{\mu\nu})}
+\frac{1}{2}\tilde{H}^{\mu\nu} (F_{\mu\nu} - 2 \partial_{\mu} A_{\nu}) \Big) 
-\int e^{-\phi}(C_3+C_1\wedge {\cal F}).  \label{eq:s2o}
\end{equation}
Here the $A_{\nu}$ equation of motion implies that
$\tilde{H}^{\mu\nu}=\epsilon^{\mu\nu\lambda}\partial_{\lambda}B$ 
for a scalar field $B$. $C_3$ and $C_1$ are determined by the condition  
\begin{equation}
d(C_3+C_1\wedge {\cal F})=d\bar{\theta}(\frac{1}{2}\psi^2+
{\cal F}\Gamma_{11})d\theta,   \label{eq:c3}
\end{equation}
with $\psi\equiv \Gamma^m \Pi_m$. Wedge products are implicit on the RHS of this 
equation and all similar subsequent equations.
Comparing the ${\cal F}$ independent terms 
in eq.~(\ref{eq:c3}),  we conclude that
\begin{equation}
dC_3+C_1 \wedge db_2 = \frac{1}{2}d\bar{\theta}\psi^2 d\theta.
\end{equation}
Eliminating the $U(1)$ gauge field in favor of the dual scalar $B$, one finds that
the dual of the action in eq.~(\ref{eq:s2o}) is 
\begin{equation}
S_D= - \int d^3 \sigma e^{-\phi} \sqrt{-{\rm det}\, G_{\mu\nu}^{\prime}}
 \label{eq:s2d}   +\int( -e^{-\phi}C_3+b_2\wedge dB) ,
\end{equation}
where
\begin{equation}
G_{\mu\nu}^{\prime} = G_{\mu\nu}+
(-e^{\phi}\partial_{\mu}B+C_{\mu})(-e^{\phi}\partial_{\nu}B+C_{\nu})
\end{equation}
and $C_{\mu}\equiv -\bar{\theta}\Gamma_{11}\partial_{\mu}\theta$. 
If we identify $B$ as the coordinate of a compact extra dimension, 
the expression appearing 
in the Born--Infeld part of the action is the standard expression 
for the induced metric of the M2-brane. 
The Wess--Zumino term also has  the appropriate structure for this
identification, since if we set $X^{11} = - e^{\phi}B$, then
$\Pi^{11}= -e^{\phi}dB+C_1 = d X^{11} + C_1$ and
\begin{eqnarray}
d(e^{-\phi}C_3-b_2 dB)&=& 
\frac{1}{2} e^{-\phi}d\bar{\theta}\Gamma_{mn}\Pi^m\Pi^n d\theta
+e^{-\phi}d\bar{\theta}\Gamma_m\Gamma_{11}\Pi^m\Pi^{11} d\theta  \\
   &= & \frac{1}{2} e^{-\phi}d\bar{\theta}\Gamma_{MN}\Pi^M\Pi^N d\theta,
\end{eqnarray} 
where $M,N$ denote 11d indices and $m,n$ denote 10d indices.
Thus eq.~(\ref{eq:s2d}) can be rewritten as 
\begin{equation}
S_D = -\int d^3 \sigma e^{-\phi} \sqrt{-{\rm det}\, G^{\prime}_{\mu\nu}} 
+\int e^{-\phi} \Omega_D , \label{eq:s2d2}
\end{equation}
where $G^{\prime}_{\mu\nu}$ and $\Omega_D$ denote 11d 
quantities. 
In order to obtain the standard M2-brane action, we should remove the 
dilaton factor. The dilaton dependence can be absorbed by the rescaling 
\begin{equation}
X^M\rightarrow e^{\frac{1}{3}\phi}X^M, \,\,\, 
\theta \rightarrow e^{\frac{1}{6}\phi}\theta . \label{eq:scaling}
\end{equation}
After this scaling, eq.~(\ref{eq:s2d2}) becomes
\begin{equation}
S_D = - \int d^3 \sigma \sqrt{-{\rm det}\, G^{11}_{\mu\nu}} 
+\int \Omega^{11}  \label{eq:s2d3} 
\end{equation}
with  $G^{11}_{\mu\nu}=\Pi^M_{\mu}\Pi^N_{\nu}\eta_{MN}$ and 
$d\Omega^{11}=-\frac{1}{2} d\bar{\theta}\Gamma_{MN}\Pi^M\Pi^N d\theta$. 
This is the standard M2-brane action~\cite{bergshoeff2}. 
Thus, as expected, we identify the M2-brane 
action (with a circular 11th dimension) as the dual of the D2-brane action. 

Let us check that the scaling that was required gives the usual relation 
between the IIA string theory and M theory. 
Comparing $G_{\mu\nu}$ appearing in eq.~(\ref{eq:s2d}) and 
$G^{11}_{\mu\nu}$ of eq.~(\ref{eq:s2d3}), 
we obtain 
\begin{equation}
G^{11}_{\mu\nu}=e^{-\frac{2}{3}\phi}G_{\mu\nu}
+e^{\frac{4}{3}\phi}(-\partial_{\mu}B+e^{-\phi}C_{\mu})
(-\partial_{\nu}B+e^{-\phi}C_{\nu}) = e^{-\frac{2}{3}\phi}G_{\mu\nu}^{\prime}
. \label{eq:g11}
\end{equation}
This correctly reproduces the relation between the 11d 
metric and the string metric in 10d~\cite{witten10}. In particular, 
 the coefficient in front of 
$(\partial B)^2$ gives the standard relation $R_{11}=e^{\frac{2}{3}\phi}$,
where $R_{11}$ is the radius of the compactified circle in the 11th  direction.  

\medskip

\section{The D3-brane}
  
The D3-brane should be self dual, {\it i.e.},  invariant under an 
$SL(2,Z)$ transformation. For the bosonic case, 
the self-duality of the D3-brane was shown in \cite{tseytlin}. So we
wish to extend the argument to the supersymmetric D3-brane action. 

Consider first the D3-brane with $e^{-\phi}=1$. The D3-brane action 
presented in \cite{aganagic} is 
\begin{equation}
S = - \int d^4 \sigma  \sqrt{-{\rm det}\, (G_{\mu\nu}+\cal F_{\mu\nu})}
+ \int (C_4+C_2\wedge {\cal F} +\frac{1}{2} C_0 F\wedge F), \label{D3action}
\end{equation}
where $C_4$ and $C_2$ are determined by the condition 
\begin{equation}
d(C_4+C_2\wedge {\cal F})=\frac{1}{6}d\bar{\theta}\tau_3\tau_1\psi^3 d \theta
+ d \bar\theta\tau_1{\cal F}\psi d\theta.  \label{eq:c4}
\end{equation}
This condition gives the useful identity 
\begin{equation}\label{eq:id3}
dC_4+C_2\wedge d{\cal F}=dC_4-C_2\wedge db_2=
\frac{1}{6}d\bar{\theta} \tau_3\tau_1 \psi^3 d\theta.
\end{equation}
The $C_0$ term in eq.~(\ref{D3action})
is absent in \cite{aganagic}, but it is a total derivative term (or boundary term)
that can be added to the action without changing the classical equations of motion. 
As in the case of the D1-brane, a constant shift of $C_0$ is a trivial classical symmetry
of the action. 

Introducing a Lagrange multiplier as before and rewriting the 
boundary term in terms of $\cal F$ instead of $F$, the action becomes 
\begin{eqnarray}
S^{\prime}&=& \int d^4 \sigma ( - \sqrt{-{\rm det}\, (G_{\mu\nu}
+\cal F_{\mu\nu})} +\frac{1}{2}\tilde{H}^{\mu\nu}(F_{\mu\nu}
-2 \partial_{\mu} A_{\nu}))  \label{eq:s3o} \\
          & & +\int (C_4+\frac{1}{2}C_0 b_2\wedge b_2 
 +(C_2+b_2 C_0)\wedge {\cal F}+\frac{1}{2}C_0
{\cal F}\wedge {\cal F}), \nonumber
\end{eqnarray}
This time the $A_{\nu}$ equation of motion is solved by 
$ \tilde{H}^{\mu\nu}  = \epsilon^{\mu\nu\lambda\sigma}\partial_{\lambda}
B_{\sigma}$.
Again, the duality transformation is similar to the bosonic case and we 
obtain
\begin{equation}
S_D =  -\int d^4 \sigma \sqrt{-{\rm det}\, \Big(G_{\mu\nu} \label{eq:s3d}
+\frac{1}{\sqrt{1+C_0^2}} (\tilde{F}_{\mu\nu}
+C_{\mu\nu}+C_0b_{\mu\nu})\Big)}
+\int \Omega_D ,
\end{equation}
where  $\tilde F = d B$ and $\Omega_D$ is given by
\begin{eqnarray}\label{eq:Omgd3}
 \Omega_D&=& C_4-b_2\wedge C_2-\frac{1}{2}C_0 b_2\wedge b_2
+b_2\wedge(\tilde{F}+C_2+C_0 b_2)  \nonumber \\ 
  & &-\frac{C_0}{2(1+C_0^2)}(\tilde{F}+C_2+C_0 b_2)
\wedge (\tilde{F}+C_2+C_0 b_2). \label{OmegaDeqn}
\end{eqnarray}

To prove that $ \Omega_D$ has 
the same form as $ \Omega $, we apply the following rotation of the
Pauli matrices:
\begin{eqnarray}\label{eq:rot}
\tau_1^{\prime}\equiv - ({\tau_3+C_0\tau_1})/{\sqrt{1+C_0^2}} \nonumber \\
\tau_3^{\prime}\equiv ({\tau_1-C_0\tau_3})/{\sqrt{1+C_0^2}} \label{rotatetau}
\end{eqnarray}
Then
\begin{equation}
\frac{1}{\sqrt{1+C_{0}^2}}
(\tilde{F}+C_2+C_0 b_2) = \tilde F^{\prime}-b_2^{\prime}
= \tilde{\cal F}^{\prime} ,
\end{equation} 
where $ \tilde F^{\prime} = \tilde F / {\sqrt{1+C_0^2}} $. From eq.~(\ref{eq:Omgd3})
\begin{equation}
d \Omega_D = d C_4 - C_2 \wedge d b_2 - 
b_2 \wedge ( d C_2 + C_0 d b_2 - 
{\sqrt{1+C_0^2}} d \tilde{\cal F}^{\prime} )  
+ ( {\sqrt{1+C_0^2}} d b_2 - C_0 
d \tilde{\cal F}^{\prime} ) \wedge \tilde{\cal F}^{\prime}  .
\end{equation}
Using eq.~(\ref{eq:rot}) one sees that the first 
parenthetical factor of the above equation vanishes, while the 
second one gives $ d\bar\theta \tau_1^{\prime}\psi d\theta $.
Using eq.~(\ref{eq:id3}) and the fact that $ \tau_3 \tau_1 = 
\tau_3^\prime \tau_1^\prime $, we finally get
\begin{equation}
d\Omega_D =\frac{1}{6}d\bar{\theta} \tau_3^{\prime}
\tau_1^{\prime} \psi^3 d\theta
+ d\bar{\theta} \tau_1^{\prime}\psi d\theta
\wedge \tilde{\cal F}^{\prime}, \label{eq:lw} 
\end{equation}
which corresponds to eq.~(\ref{eq:c4}). Thus the dual action can be rewritten as 
\begin{equation}\label{equ:S3d}
S_D = -\int d^4 \sigma ( \sqrt{-{\rm det}\, (G_{\mu\nu}
+\tilde{\cal F}_{\mu\nu}^{\prime})}
+\int (C_4^{\prime} + C_2^\prime \wedge 
\tilde {\cal F}^{\prime} - \frac{1}{2} 
C_0 \tilde F^{\prime} \wedge \tilde F^{\prime}).
\end{equation}
This action can be interpreted as a D3-brane in the presence of both
constant dilaton and axion backgrounds. However, to make this identification,
we must present the general formula with such backgrounds.
 
In the string metric, the action including arbitrary constant 
dilaton and axion backgrounds is
\begin{equation}
S = - \int d^4 \sigma e^{-\phi} \sqrt{-{\rm det}\, (G_{\mu\nu}+\cal F_{\mu\nu})}
+\int e^{-\phi} (C_4+C_2\wedge {\cal F} ).
\end{equation}
In order to get to the Einstein metric, which is invariant under $SL(2,R)$ transformations, 
we rescale
\begin{equation}
X^m \rightarrow e^{\phi/4} X^m \quad {\rm and} \quad
\theta \rightarrow e^{\phi/8} \theta.
\end{equation} 
The action becomes
\begin{equation}
S = -\int d^4 \sigma \sqrt{-{\rm det}\, (G_{\mu\nu}
+e^{-\frac{\phi}{2}} F_{\mu\nu}-b_{\mu\nu})}
+\int \Big( C_4 + C_2 \wedge (e^{-\frac{\phi}{2}} F - b_2) \Big).
\end{equation}
We now add a Lagrange multiplier term $ \frac{1}{2} \tilde H^{\mu\nu} (F_{\mu\nu}
- 2\partial_{\mu} A_{\nu}) $ and a boundary term $ \frac{1}{2} C_0 F \wedge F $.
If we define $ F^\prime \equiv e^{-\frac{\phi}{2}} F $, $ \tilde H^{\prime} = 
e^{\frac{\phi}{2}} \tilde H $, and $ C_0^\prime = e^\phi C_0 $, 
the action expressed in terms
of primed quantities is just (\ref{eq:s3o}), so we can read off the dual action 
from eqs.~(\ref{eq:s3d}) and (\ref{OmegaDeqn}).
The resulting action is
\begin{equation} 
S_D=  - \int d^4 \sigma  \sqrt{-{\rm det}\, (G_{\mu\nu}
+\frac{1}{\sqrt{1+e^{2\phi}C_0^{ 2}}} (e^{\frac{\phi}{2}}
\tilde{F}_{\mu\nu}
+C_{\mu\nu}+e^{\phi}C_0 b_{\mu\nu})}
+\int \Omega_D,  \label{eq:s3d2}
\end{equation}
where
\begin{eqnarray}
 \Omega_D&=& C_4-b_2\wedge C_2-\frac{1}{2}e^{\phi}C_0  
b_2\wedge b_2
+b_2\wedge(e^{\frac{\phi}{2}}\tilde{F} +C_2+e^{\phi}C_0  b_2)
  \nonumber  \label{eq:lw2} \\ 
 & &-\frac{e^{\phi}C_0 }{2(1+e^{2\phi}C_0^{ 2})}
(e^{\frac{\phi}{2}}\tilde{F} +C_2+e^{\phi}C_0  b_2)
\wedge (e^{\frac{\phi}{2}}\tilde{F} +C_2+e^{\phi}C_0  b_2).
\end{eqnarray}
The kappa symmetry of this action follows from that of eq.~(\ref{eq:s3d}). 
Also, we can check the transformation of the dilaton and 
the axion under the duality transformation. 
{}From the coefficient of $\tilde{F}$ in the Born--Infeld part of 
eq.~(\ref{eq:s3d2}), we obtain the transformation
\begin{equation}
e^{-\phi} \rightarrow \frac{e^{\phi}}{1+e^{2\phi}C_0^{2}}
=\frac{1}{e^{\phi}+e^{-\phi}C_0^{2}}, 
\end{equation}
and from the coefficient of $\tilde{F} \wedge \tilde{F}$  
we have 
\begin{equation}
C_0 \rightarrow -\frac{e^{2\phi}C_0}
{1+e^{2\phi}C_0^{2}}=-\frac{e^{\phi}C_0 }
{e^{-\phi}+e^{\phi}C_0^{2}}.
\end{equation}
Thus, the dilaton and the axion undergo the expected
$SL(2,Z)$ transformation. 
Combining this symmetry with the symmetry under a 
constant shift of $C_0$, one deduces that the D3-brane 
action has $SL(2,R)$ symmetry classically. Of course, this is reduced to 
$SL(2,Z)$  by quantum effects. 

\medskip

\section{The D4-brane}

The D4-brane action plus a Lagrange multiplier term is given by 
\[
S = \int d^5 \sigma (  - \sqrt{-{\rm det}\, (G_{\mu\nu}+\cal F_{\mu\nu})}
+\frac{1}{2}\tilde{H}^{\mu\nu} (F_{\mu\nu} - 2 \partial_{\mu} A_{\nu})) 
\]
\begin{equation}
- \int (C_5+C_3\wedge {\cal F} +\frac{1}{2} C_1 \wedge {\cal F}\wedge {\cal F}).
\end{equation}
The $A_{\nu}$ equation of motion implies
that $\tilde{H}^{\mu\nu}=\frac{1}{6}\epsilon^{\mu\nu\lambda\sigma\tau}
H_{\lambda\sigma\tau}$ with $H=dB$. Also $C_1=\bar{\theta}\Gamma^{11}d\theta$
and 
\begin{eqnarray}
C_3&=&\frac{1}{2}\bar{\theta}\Gamma_{m_1m_2}d\theta(dX^{m_1}dX^{m_2}
+\bar{\theta}\Gamma^{m_1}d\theta dX^{m_2}
+\frac{1}{3}\bar{\theta}\Gamma^{m_1}d\theta 
\bar{\theta}\Gamma^{m_2}d\theta) \\
  & &+ \frac{1}{2}\bar{\theta}\Gamma_{11}\Gamma_{m_1}d\theta
\bar{\theta}\Gamma_{11}d\theta  
(dX^{m_1}+\frac{2}{3}\bar{\theta}\Gamma^{m_1}d\theta),
\end{eqnarray}
while $C_5$ is determined by 
\begin{equation}
dC_5=\frac{1}{24}d\bar{\theta} \Gamma^{11} \psi^4 d\theta+db_2\wedge C_3.
\end{equation}

The action $S$ can be written in two parts 
$S =S_1+S_2$, where 
\begin{equation}
S_1= - \int d^5 \sigma \sqrt{-{\rm det}\, (G_{\mu\nu}+\cal F_{\mu\nu})}
+\int ({\cal H}\wedge {\cal F} - \frac{1}{2}C_1 \wedge {\cal F}\wedge {\cal F}),
\end{equation}
\begin{equation}\label{equ:S4d2}
S_2 \equiv \int \Omega  =\int ( - C_5+H\wedge b_2),
\end{equation}
and ${\cal H}\equiv H - C_3$. 
The appendix  shows that after the duality transformation 
one obtains the dual action $S_D = S_{1D} + S_{2}$, where 
\begin{equation}\label{equ:S4d1}
S_{1D} = - \int d^5 \sigma \left(\sqrt{-G}\sqrt{1+z_1+\frac{z_1^2}{2}-z_2}
+\frac{\epsilon_{\mu\nu\lambda\sigma\tau}
C^{\mu}\tilde{{\cal H}}^{\nu\lambda}
\tilde{{\cal H}}^{\sigma\tau}}{8(1+C_1^2)}\right).
\end{equation}
Here 
\begin{equation}
z_1\equiv \frac{{\rm tr}(\tilde{G}\tilde{{\cal H}}\tilde{G}\tilde{{\cal H}})}
{2(-G)(1+C_1^2)} \quad
 {\rm and } \quad
z_2\equiv \frac{{\rm tr}(\tilde{G}\tilde{{\cal H}}\tilde{G}\tilde{{\cal H}}\tilde{G}
\tilde{{\cal H}}\tilde{G}\tilde{{\cal H}})}{4(-G)^2(1+C_1^2)^2},
\end{equation} 
where $\tilde{G}_{\mu\nu}\equiv G_{\mu\nu}+C_{\mu}C_{\nu}$ and 
$\tilde{{\cal H}}^{\mu\nu}=\frac{1}{6}\epsilon^{\mu\nu\lambda\sigma\tau}
{\cal H}_{\lambda\sigma\tau}$. 
The duality transformation leaves $S_2 $ unchanged. The Wess--Zumino 
term $\Omega$ is given by 
\begin{equation}
d\Omega =dC_5+db_2\wedge H
= - \frac{1}{24}d\bar{\theta} \Gamma^{11} \psi^4 d\theta-d\bar{\theta}
\Gamma_{11}\psi d\theta\wedge {\cal H}.
\end{equation}
This dual action D4-brane action is identical to the action obtained by
double-dimensional reduction of the M5-brane,
which was given in sect. 6 of ref.~\cite{APPS}.   In that work, the radius of the
compact dimension was set equal to one, which corresponds to setting the
IIA dilaton equal to zero, as done here.

Let us now consider how the analysis described above generalizes
when a constant dilaton background field is included. 
In this case, the D4-brane action is
\begin{equation}
S^{\prime }= - \int d^5 \sigma  e^{-\phi}\sqrt{-{\rm det}\, (G_{\mu\nu}
+\cal F_{\mu\nu})}
- \int e^{-\phi}(C_5+C_3\wedge {\cal F} 
+\frac{1}{2} C_1 \wedge {\cal F}\wedge {\cal F}).
\end{equation}
The action after the duality transformation is 
\begin{equation}
S^{\prime}= - \int d^5 \sigma \left( e^{-\phi}   \label{eq:sd43}
\sqrt{-G}\sqrt{1+e^{2\phi}z_1+e^{4\phi}(\frac{z_1^2}{2}-z_2)}
+\frac{\epsilon_{\mu\nu\lambda\sigma\tau}
e^{\phi}C^{\mu}\tilde{{\cal H}}^{\nu\lambda}
\tilde{{\cal H}}^{\sigma\tau}}{8(1+C_1^2)}\right)+\int \Omega^{\prime},
\end{equation}
Here $z_1$ and $z_2$ are defined as before, but now
\begin{equation} 
{\cal H}\equiv H - e^{-\phi}C_3   \label{eq:h2}
\end{equation}
 and $\Omega^{\prime}$ is 
determined by the equation 
\begin{equation}
d\Omega^{\prime}
= - \frac{1}{24}e^{-\phi}d\bar{\theta} \Gamma_{11} \psi^4 d\theta-d\bar{\theta}
 \Gamma_{11}\psi d\theta\wedge {\cal H}. 
\end{equation}

It now remains to  show that this action agrees with the one obtained by
double dimensional reduction of the M5-brane, when the 
analysis of ref.~\cite{APPS} is generalized to include the radius of the
compact dimension and one identifies that radius with ${\rm exp} (2\phi/3)$.
This means that it should coincide with the double dimensional 
reduction of the M5-brane using the 11d metric $G^{11}_{\mu\nu}$ introduced
in eq.~(\ref{eq:g11})
in connection with the D2-brane. In eq.~(\ref{eq:g11})
\begin{equation}
G^{11}_{\mu\nu}=e^{-\frac{2\phi}{3}}G^{\prime}_{\mu\nu}
=e^{-\frac{2\phi}{3}}\Pi^{M}_{\mu}\Pi^{N}_{\nu}\eta_{MN},
\end{equation}
where $\Pi^{11}=dX^{11}+C_1=-e^{\phi}dB+C_1$. 
In ref.~\cite{APPS}, $G^{11}_{\mu\nu}$ is the pullback of the 11d 
flat metric. In order to compare the M5-brane action 
with the dual 4-brane action, 
in which the string metric is the usual flat metric, we need to rescale 
the variables appropriately. The required scaling is 
\begin{equation} 
X^M \rightarrow e^{-\frac{1}{3}\phi}X^M, \, \,\, \theta \rightarrow  
e^{-\frac{1}{6}\phi}\theta.  
\end{equation}
This is just the inverse of the transformation in eq.~(\ref{eq:scaling}),
which was used to convert to the 11d canonical flat metric from the string metric. 
Since  $X^{11}$ is defined to be $-e^{\phi}B $ in eq.~(\ref{eq:g11}), 
after this scaling it becomes $X^{11}=-e^{\frac{2}{3}\phi}B$.
Carrying out the double dimensional reduction by setting 
$B= - \sigma^5$ in eq.~(\ref{eq:g11})\footnote{The circular 11th
dimension has 
circumference $2\pi R_{11}$, and $B$ runs between 0 and $2\pi$,
so $R_{11} \sim {\rm exp} (2\phi/3)$.}
and dropping the $\sigma^5$ dependence 
of the other variables,  we obtain 
\begin{equation}
G^{11}_{\hat\mu\hat\nu}=\left( \begin{array}{cc}
 e^{-\frac{2}{3}\phi}G_{\mu\nu}+e^{-\frac{2}{3}\phi}C_{\mu}C_{\nu} 
&  e^{\frac{1}{3}\phi}C_{\mu} \\ & \\
e^{\frac{1}{3}\phi}C_{\nu} & e^{\frac{4}{3}\phi}
 \end{array}    \right).   \label{eq:gdd}
\end{equation}
Here $\hat\mu, \hat\nu$ run from 0 to 5 and $\mu, \nu$ run from 0 to 4 
and $G_{\mu\nu}=\Pi^m_{\mu}\Pi^n_{\nu}\eta_{mn}$. The rescaling 
also gives $C_3 \rightarrow e^{-\phi}C_3$, which implies that the 
quantity ${\cal H}$ 
that is used in the double dimensional reduction of M5-brane 
is the same as in eq.~(\ref{eq:h2}).  Thus we can conclude that
the double dimensional reduction of the M5-brane with these rescaled variables
gives the same action as the dual 4-brane action 
with a constant dilaton in eq.~(\ref{eq:sd43}).  

\medskip

\section{Duality Transformations of Gauge-Fixed Theories}

The analysis can be repeated for the gauge-fixed D-branes 
of~\cite{aganagic}. As shown there, one can go to a static gauge 
by imposing $ X^\mu = \sigma^\mu $ for $\mu = 0, \ldots, p$ and 
setting $ \theta_2 = 0 $ ($ \theta_1 = 0 $) for IIA (IIB), respectively. 
If  we do not include any background fields, the Wess--Zumino term vanishes in this gauge.
Denoting the component of the spinor that survives by $\lambda$
and the transverse components of $X^{\mu}$ by $\phi^i$ with $i= p+1, \ldots, 9$, we 
get the action~\cite{aganagic}
\begin{equation}
S = - \int d^{p+1} \sigma \sqrt{- {\rm det}\,  (G_{\mu\nu} + {\cal F}_{\mu\nu})}
\end{equation} 
with
\begin{equation}
G_{\mu\nu} = \eta_{\mu\nu} + \partial_\mu \phi^i \partial_\nu \phi^i 
- 2\bar\lambda (\Gamma_{(\mu} 
+ \Gamma_i \partial_{(\mu} \phi^i) \partial_{\nu)} \lambda 
+ \bar\lambda \Gamma^m \partial_\mu \lambda \bar\lambda 
\Gamma_m \partial_\nu \lambda ,
\end{equation}
\begin{equation}
{\cal F}_{\mu\nu} = F_{\mu\nu} - b_{\mu\nu} = F_{\mu\nu} 
-  2\bar\lambda (\Gamma_{[\mu} 
+ \Gamma_i \partial_{[\mu} \phi^i) \partial_{\nu]} \lambda.
\end{equation}
Since the Wess--Zumino term vanishes in this gauge, the dual 
actions have the same form as in eq.~(\ref{moregen}).

The IIA cases are straightforward in this picture: the dual action 
corresponds to the dual theory in the same kind of static gauge 
($ X^\mu = \sigma^\mu $, $ \theta_2 = 0 $). Indeed, for $ p = 2 $
\begin{equation}
S_D = \int d^{3} \sigma \left( - \sqrt{- {\rm det}\, (G_{\mu\nu} 
+ \partial_\mu B \partial_\nu B)}
	+ {1 \over 2} \epsilon^{\mu\nu\rho} b_{\mu\nu} \partial_\rho B \right).
\end{equation}
This is precisely the action (\ref{eq:s2d}) in the static gauge, 
because for $ \theta_2 = 0 $ both $C_1$ and $C_3$ vanish.
Similarly, for $ p = 4 $
\begin{equation}
S_D =  \int d^{5}\sigma 
  \left( -  \sqrt{ - G ( 1 + z_1 + {z_1^2 \over 2} - z_2 ) } + 
{1 \over 2} \tilde H^{\mu\nu} b_{\mu\nu} \right),
\end{equation}
where the $z$'s are similar to the bosonic ones, involving only $G$ and 
$\tilde H$. 
This corresponds to eqs.~(\ref{equ:S4d2}) and (\ref{equ:S4d1}) for 
$ C_1 = C_3 = C_5 = 0 $.

The IIB duals are a bit puzzling at first sight. For $ p = 1 $
\begin{equation}
S_D =  \int d^{2}  \sigma \left( - \sqrt{ 1 + \Lambda^2 } 
\sqrt{- {\rm det}\, G_{\mu\nu} }
	+ {1 \over 2}\Lambda \epsilon^{\mu\nu} b_{\mu\nu} \right)
\end{equation}
and for $ p = 3 $
\begin{equation}
S_D =  \int d^{4} \sigma  \left( -  \sqrt{- {\rm det}\, (G_{\mu\nu} 
+ \tilde F_{\mu\nu})}
	+ {1 \over 4} \epsilon^{\mu\nu\rho\tau} 
\tilde F_{\mu\nu} b_{\rho\tau}\right) .
\end{equation}
The 3-brane is supposed to be self-dual, yet the dual theory 
looks different: it has $F$ instead of ${\cal F}$ under 
the square root and nonvanishing Wess--Zumino term. 
Also, the two terms in the dual 1-brane have different coefficients, 
unlike the fundamental string. 
The explanation is that in the IIB case 
the dual theories correspond to the $ \theta_1^\prime = 0 $ gauge, 
where the prime means that we first undo the rotation in
$\tau$ space (see eq.~(\ref{rotatetau}))
and then impose $ \theta_1
 = 0 $. For the 3-brane, this amounts to imposing the gauge 
$\theta_1 = \theta_2 \equiv \lambda / \sqrt{2} $ .
In this gauge $b^\prime_2$ vanishes, but the Wess--Zumino term contributes. 
In fact, the above formulas agree 
with eqs.~(\ref{eq:S1d}) and (\ref{equ:S3d}) in the 
$ \theta_1^\prime = 0 $ gauge.

\medskip

\section{Discussion}

We have explored the duality transformation
properties of  super Dp-branes for $p= 1,2,3,4$. In each case, the results 
agreed with the expectations suggested by standard dualities.
For the D5-brane and  higher-dimensional objects, we have not yet been able 
to carry out the analysis. As explained in sect. 2,
it is much more difficult to write 
the Born--Infeld action in terms of the dual gauge field for $p\geq 5$ even in the 
bosonic case. For example, the dual D5-brane, which ought to correspond to
the solitonic 5-brane of the IIB theory, would be expressed in term of a
world-volume 3-form potential. Perhaps a more powerful approach is
required to make this problem tractable.

For the most part, our analysis has been classical and limited to
flat backgrounds. The results should not depend on these restrictions, however.
 
\newpage
\section*{Appendix  -- Duality Transformation of D4-brane }

The following D4-brane action appeared in sect. 6
\begin{equation}
S_1= - \int d^5 \sigma (  \sqrt{-{\rm det}\, (G_{\mu\nu}+\cal F_{\mu\nu})}
+\int ({\cal H}\wedge {\cal F} - \frac{1}{2}C_1 \wedge{\cal F}\wedge {\cal F}). 
\label{eq:s4o}
\end{equation}
Because of the Wess-Zumino term, the duality transformation is considerably
more complicated in this case than for the bosonic truncation.
Because of the general covariance, it is sufficient to consider the flat 
limit $G_{\mu\nu} =\eta_{\mu\nu}$. The $G$ dependence is easily reinstated in the
answer. Also, we can use the Lorentz invariance of this flat limit
to choose a special basis where the only nonzero 
components of ${\cal F}_{\mu\nu}$ and $C_{\mu}$ are
\begin{equation}
{\cal F}_{12}=-{\cal F}_{21}=f_1, \,\,\, {\cal F}_{34}=-{\cal F}_{43}=f_2 
\end{equation}
\begin{equation}
C_{\mu}=(c_0, c_1,0,c_3,0), \nonumber
\end{equation}
where we use lower case $c$'s, because numerical subscripts on $C$'s denote
differential forms.
{}From the equations of motion following from eq.~(\ref{eq:s4o}), we then obtain 
the following nonzero components of $\tilde{{\cal H}}^{\mu\nu}$
\begin{equation} 
 \tilde{{\cal H}}^{02}=-\tilde{{\cal H}}^{20}= - c_1 f_2, \,\,\, \label{eq:h}
\tilde{{\cal H}}^{04}=-\tilde{{\cal H}}^{40}= - c_3 f_1
\end{equation}
\begin{equation}
\tilde{{\cal H}}^{12}=-\tilde{{\cal H}}^{21}
= \sqrt{\frac{1+f_2^2}{1+f_1^2}}f_1 + c_0 f_2, 
\nonumber
\end{equation}
\begin{equation}
\tilde{{\cal H}}^{34}=-\tilde{{\cal H}}^{43}
= \sqrt{\frac{1+f_1^2}{1+f_2^2}}f_2
 + c_0 f_1. \nonumber
\end{equation}
It is useful to define $y^{\nu}\equiv C_{\mu}\tilde{{\cal H}}^{\mu\nu}$,
whose nonzero components are 
\begin{equation}
y^2=   \sqrt{\frac{1+f_2^2}{1+f_1^2}}c_1 f_1, \quad  \label{eq:y}
y^4=   \sqrt{\frac{1+f_1^2}{1+f_2^2}}c_3 f_2.
\end{equation}

If $\tilde{\cal H}^{02}=\tilde{\cal H}^{04}=0$, or equivalently $y^{\mu}=0$, 
the analysis would be very similar to the D3-brane with nonzero $C_0$. 
The dual action of eq.~(\ref{eq:s4o}) in this case is 
\begin{equation}
S_{1D}= - \int d^5 \sigma \left(\sqrt{-G}
\sqrt{1+\frac{\tilde{{\cal H}}^2}{2(-G)(1+C_1^2)}
+\frac{(\tilde{{\cal H}}^2)^2-2 \tilde{{\cal H}}^4}{8(-G)^2 (1+C_1^2)^2}} 
+\frac{\epsilon_{\mu\nu\lambda\sigma\tau}
C^{\mu}\tilde{{\cal H}}^{\nu\lambda}\tilde{{\cal H}}^{\sigma\tau}}
{8(1+C_1^2)}\right), \label{eq:s4d1}
\end{equation}
where $\tilde{{\cal H}}^2 = {\rm tr} (G\tilde{\cal H} G \tilde{\cal H})$ and
similarly for $\tilde{{\cal H}}^4$.
When $y^{\mu}$ is nonzero, the analysis becomes more complicated. One could try 
to rewrite the action (\ref{eq:s4o}) in terms of $\tilde{\cal H}$ using the 
Lorentz invariant quantities made out of $\tilde{{\cal H}}^{\mu\nu}$ and 
$y^{\mu}$, and 
using the relation between ${\cal F}$ and $\tilde{{\cal H}}$ obtained 
from the equations of motion. 
Instead of doing that, we will take advantage of the fact that we already know the
answer (from double-dimensional reduction of the M5-brane action). 
Defining $\tilde{G}_{\mu\nu}\equiv\eta_{\mu\nu}+C_{\mu}C_{\nu}$ and  
\begin{equation}
z_1\equiv\frac{{\rm tr}(\tilde{G}\tilde{{\cal H}}\tilde{G}\tilde{{\cal H}})}
{2(-G)(1+C_1^2)}   
=\frac{\tilde{{\cal H}}^{\mu\nu}\tilde{{\cal H}}_{\nu\mu}-2y^{\mu} y_{\mu}}
{2(1+C_1^2)}, \label{eq:z1}
\end{equation}
\begin{equation}
z_2\equiv \frac{{\rm tr}(\tilde{G}\tilde{{\cal H}}\tilde{G}\tilde{{\cal H}}\tilde{G}
\tilde{{\cal H}}\tilde{G}\tilde{{\cal H}})}{4(-G)^2(1+C_1^2)^2}
=\frac{\tilde{{\cal H}}^4-4y^{\mu}\tilde{{\cal H}}_{\mu\alpha}
\tilde{{\cal H}}^{\alpha\lambda}
y_{\lambda}+2(y^{\mu} y_{\mu})^2}{4(1+C_1^2)^2} ,   \label{eq:z2}
\end{equation}
we consider the expression
\begin{equation}
\sqrt{1+z_1+\frac{z_1^2}{2}-z_2}
+\frac{\epsilon_{\mu\nu\lambda\sigma\tau}
C^{\mu}\tilde{{\cal H}}^{\nu\lambda}\tilde{{\cal H}}^{\sigma\tau}}{8(1+C_1^2)}.
\end{equation}
It is a matter of calculation to show that this is equal to 
\begin{equation}
\sqrt{-{\rm det}\, (\eta_{\mu\nu}+\cal F_{\mu\nu})}
- \frac{1}{2}\tilde{{\cal H}}^{\mu\nu}{\cal F}_{\mu\nu}
+\frac{1}{8}\epsilon^{\mu\nu\lambda\sigma\tau}
C_{\mu}{\cal F}_{\nu\lambda}{\cal F}_{\sigma\tau}
\end{equation}
using the form of $\tilde{{\cal H}}^{\mu\nu}$ and $y^{\nu}$ of 
eqs.~(\ref{eq:h}) and 
(\ref{eq:y}) in the special basis.\footnote{For example, one can check 
\begin{eqnarray}
\sqrt{1+z_1+\frac{z_1^2}{2}-z_2}&=& 
\frac{1-f_1^2f_2^2+c_1^2(1+f_2^2)+c_3^2(1+f_1^2)}
{(1+C^2)\sqrt{(1+f_1^2)(1+f_2^2)}}   \nonumber         \\
   & & -\frac{c_0^2(1+f_1^2)(1+f_2^2)+2c_0 f_1 f_2\sqrt{(1+f_1^2)(1+f_2^2)}}
{(1+C^2)\sqrt{(1+f_1^2)(1+f_2^2)}}.          \nonumber
\end{eqnarray}
} 
Now putting back the metric dependence, we conclude that in the general case
\begin{equation}
S_{1D}= - \int d^5 \sigma \left(\sqrt{-G}\sqrt{1+z_1+\frac{z_1^2}{2}-z_2}
+\frac{\epsilon_{\mu\nu\lambda\sigma\tau}
C^{\mu}\tilde{\cal H}^{\nu\lambda}\tilde{\cal H}^{\sigma\tau}}{8(1+C_1^2)}\right),
\end{equation}
where $z_1$ and $z_2$ are defined as in eqs.~(\ref{eq:z1}) and (\ref{eq:z2}) with 
$\tilde{G}_{\mu\nu}=G_{\mu\nu}+C_{\mu}C_{\nu}$.

\end{document}